\documentclass[10pt]{sig-alternate-10pt-sigcomm}
\usepackage{amsmath}
\usepackage{mathptmx}
\usepackage{mathtools}
\usepackage{color}
\usepackage{amsmath}
\usepackage{cleveref}
\usepackage{makecell}
\usepackage{graphicx}
\usepackage{url}
\usepackage{float}
\usepackage[ruled,vlined,linesnumbered]{algorithm2e}
\usepackage[caption = false]{subfig}
\usepackage{capt-of}
\usepackage{xspace}
\usepackage{amssymb}
\usepackage{booktabs}
\newcommand{\ra}[1]{\renewcommand{\arraystretch}{#1}}

\usepackage{tabularx}
\usepackage{multirow}

\usepackage{listings}
\usepackage[utf8]{inputenc}
\usepackage{color}
\usepackage{mathtools}
\DeclarePairedDelimiter{\ceil}{\lceil}{\rceil}
\usepackage{comment}

\definecolor{dkgreen}{rgb}{0,0.6,0}
\definecolor{gray}{rgb}{0.5,0.5,0.5}
\definecolor{mauve}{rgb}{0.58,0,0.82}

\newcommand{\XYZs}{Dapper's\xspace}
\newcommand{\XYZ}{Dapper\xspace}
\newcommand{\cwnd}[1]{$inf{\_}cwnd$}

\newcommand{\theo}[1]{\textcolor{blue}{(Theo: #1)}}

\usepackage{etoolbox}
\makeatletter
\patchcmd{\maketitle}{\@copyrightspace}{}{}{}
\makeatother

\begin{document}
\title{Dapper: Data Plane Performance Diagnosis of TCP}

\numberofauthors{3}
\author{
\alignauthor 
Mojgan Ghasemi\\
\affaddr{Princeton University}
\email{mojgan@cs.princeton.edu}    
\alignauthor
Theophilus Benson\\
\affaddr{Duke University}
\email{tbenson@cs.duke.edu}    
\alignauthor
Jennifer Rexford\\
\affaddr{Princeton University}
\email{jrex@cs.princeton.edu}     
}

\maketitle
\thispagestyle{empty}

\begin{abstract}
With more applications moving to the cloud, cloud providers need to diagnose performance problems in a timely manner. 
Offline processing of logs is slow and inefficient, and instrumenting the end-host network stack would violate the tenants' rights to manage their own virtual machines (VMs).  
Instead, our \XYZ system analyzes TCP performance in real time \emph{near} the end-hosts (e.g., at the hypervisor, NIC, or top-of-rack switch). \XYZ determines whether a connection is limited by the sender (e.g., a slow server competing for shared resources), the network (e.g., congestion), or the receiver (e.g., small receive buffer). 
Emerging edge devices now offer flexible packet processing at high speed on commodity hardware, making it possible to monitor TCP performance in the data plane, at line rate. We use P4 to prototype \XYZ and evaluate our design on real and synthetic traffic.  To reduce the data-plane state requirements, we perform lightweight \emph{detection} for all connections, followed by heavier-weight \emph{diagnosis} just for the troubled connections.

\end{abstract}
\section{Introduction}
\label{s:intro}
Public clouds need timely and accurate visibility into application performance to ensure high utilization (to reduce cost) and good performance (to satisfy tenants)~\cite{INT}.  Yet, cloud providers must monitor performance within their own infrastructure, since they
cannot modify the end-host network stack without violating tenant control over their own VMs. 
Studies reveal that more than 99\% of traffic in some data centers is TCP traffic~\cite{Alizadeh:2010:DCT:1851182.1851192}. However, TCP measurements collected in the network core would not have an end-host perspective (e.g., of round-trip times) and would require combining data from multiple switches because the two directions of a connection may traverse different paths.

\begin{figure}
\centering
\subfloat{\includegraphics[height = 1in]{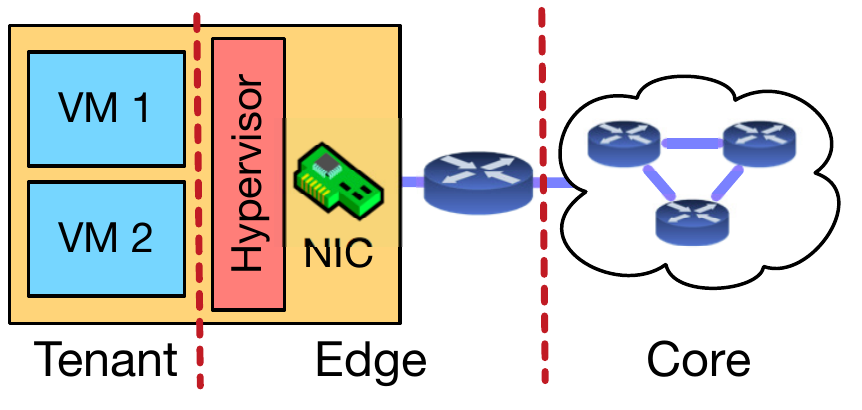}}
\caption{\XYZ monitors performance at network edge}
\label{fig:rtd_measure}
\end{figure}

Instead, measurement at the ``edge''---in the hypervisor, NIC, or top-of-rack switch, as shown in Figure~\ref{fig:rtd_measure}---offers the best viable alternative.  The edge (i) sees all of a TCP connection's packets in both directions, (ii) can closely observe the application's interactions with the network without the tenant's cooperation, and (iii) can measure end-to-end metrics from the end-host perspective (e.g., path loss rate) because it is only one hop away from the end-host. 

Fortunately, emerging edge devices offer flexible packet processing at line rate, in software switches~\cite{Shahbaz:2016:PPP:2934872.2934886}, NICs~\cite{P4workshop-NIC}, and hardware switches~\cite{RMT} that are programmable using languages like P4~\cite{Bosshart:2014:PPP:2656877.2656890}. 
New capabilities, such as flexible parsing and registers that maintain state open up the possibility of detecting and diagnosing 
TCP performance problems directly in the data plane.
A major challenge for data-plane connection diagnosis is finding a ``sweet spot'' that balances the need for \emph{fine-grained} metrics for diagnosis, while remaining \emph{lightweight} enough to run across a wide range of devices with limited capabilities.

\begin{figure}
\centering
\subfloat{\includegraphics[height=1.4in]{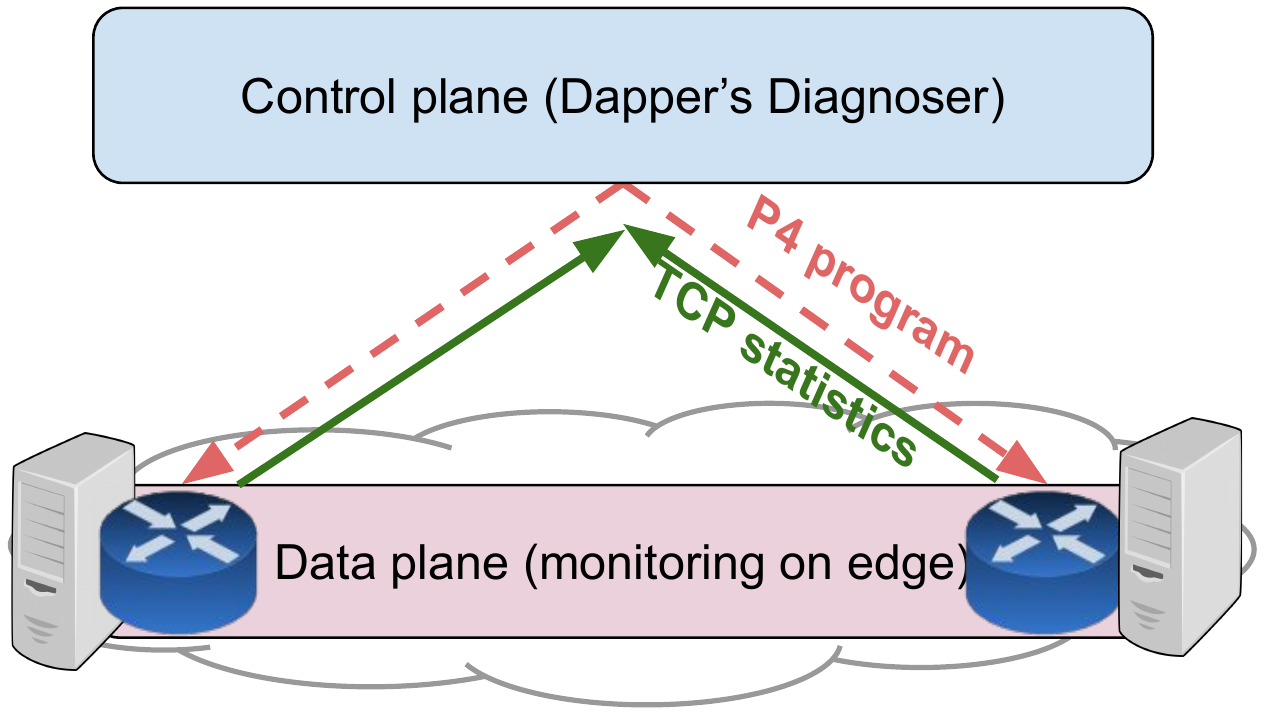}}
\caption{\XYZ's architecture has two parts: (1) data plane monitoring on edge via P4, (2) diagnosis techniques in control plane based on the collected TCP statistics }
\label{fig:architecture}
\end{figure}

This paper presents \XYZ, a \textbf{Da}ta-\textbf{P}lane \textbf{Per}formance diagnosis tool that infers TCP bottlenecks by analyzing packets in real time at the network edge, as shown in Figure~\ref{fig:architecture}. \XYZ quantifies the contribution of the sender, network, and receiver to poor end-to-end performance. Table~\ref{table:problems_list} shows examples of problems that can limit a TCP connection's performance. Identifying the entity responsible for poor performance is often the most time-consuming and expensive part of failure detection and can take from an hour to days in data centers~\cite{Arzani:2016:TBG:2934872.2934884}.  Once the bottleneck is correctly identified, specialized tools within that component can pinpoint the root cause. To achieve this goal, we need to infer essential TCP metrics.  Some of them are easy to infer (e.g., counting number of bytes or packets sent or received), while others are more challenging (e.g., congestion and receive windows). 

\XYZ analyzes header fields, packet sizes, timing, and the relative spacing of data and ACK packets, to infer the TCP state and the congestion and receive window sizes. A unique challenge in a public IaaS clouds is that different tenants run different versions of TCP, each possibly with tuned parameters. Thus, our techniques must be applicable to a heterogeneous set of TCP connections. The advantage of a data-plane performance diagnosis, apart from line-rate diagnosis, is that the data plane can use this information for quick decision making (e.g., load balancing for network-limited connections). However, a data-plane monitoring tool often has more resource constraints: limited state and limited number of arithmetic and Boolean operations  per packet. We discuss design challenges and the necessary steps for \XYZ to run in the data plane, including reducing the accuracy of some measurements (e.g., RTT) to lower the amount of per-flow state, as well as two-phase monitoring where we switch from collecting \emph{lightweight} metrics for \emph{all} flows to only collecting \emph{heavyweight} metrics for \emph{troubled} ones.

\textbf{Roadmap:} Section~\ref{s:infer} explains the TCP performance bottlenecks we identify and how to infer the metrics necessary to detect them. Section~\ref{s:diagnosis} explains how \XYZ diagnoses performance problems from the inferred statistics. Section~\ref{s:hardware} explains how to monitor TCP connections in real time using commodity packet processors programmed using P4~\cite{Bosshart:2014:PPP:2656877.2656890,p4spec}. Section~\ref{s:mid} discusses our two-phase monitoring to reduce the memory overhead in the data plane.  Section~\ref{s:eval} evaluates the overhead and performance of our system.  Section~\ref{s:relatedwork} discusses related work, and Section~\ref{s:concl} concludes the paper.

\section{TCP Performance Monitoring}
\label{s:infer}
A TCP connection may have bottlenecks at the sender, receiver, or the network. In Table~\ref{table:problems_list}, we present several examples of performance problems that may arise at each location.  With many applications and TCP variants in a public cloud, it is challenging to decide what minimal set of metrics to collect that are both affordable (i.e., does not consume a lot of resources) and meaningful (i.e., helps in diagnosis). 
In this section, we discuss the metrics we collect to diagnose the performance bottlenecks at each component, and the streaming algorithms we use to infer them.

\begin{table}[t]
\ra{1.3}
\scriptsize
\begin{tabular}{p{1cm}p{7cm}}
\toprule 
\textbf{Location} & \textbf{Performance Problems} \\
\toprule 
Sender & slow data generation rate due to resource constraints, not enough data to send (non-backlogged)\\
\midrule 
Network & congestion (high loss and latency), routing changes, limited bandwidth \\
\midrule 
Receiver & delayed ACK, small receive buffer\\
\bottomrule 
\end{tabular}
\caption{TCP performance problems at each component}
\label{table:problems_list}
\end{table}

We denote a TCP connection by a bi-directional 4-tuple of server and client IP addresses and port numbers. We focus on the performance of the data transmission from the server, where the server sends data and receives ACKs, as shown in Figure~\ref{fig:handshake}. Hence, we monitor traffic close to the server, for several reasons.
First, being close to the server enables us to imitate the internal state of the server's congestion control and monitor how quickly the server responds to ACKs. Second, if a connection is end-user facing (as opposed to east-west traffic in the data center), we do not have access to the client's edge, so would need to monitor near the server. Finally, monitoring at one end reduces the overhead and avoids redundancy as opposed to keeping per-flow state at both ends. 
Note that for east-west traffic, a cloud provider can enable monitoring at both ends of a connection, which would offer higher accuracy and better visibility into the connection's state by collecting more statistics on the client-side (e.g., delay in sending ACKs).


\subsection{Inferring Sender Statistics}
Performance problems at the sender limit the TCP connection's performance; for example, an application that is constrained by its host's resources (e.g., slow disk or limited CPU in the VM) may generate data at a lower rate than the network or receiver can accept, or it simply may not have more data to send; this application is referred to as \emph{non-backlogged}.
To find such problems, we measure the \emph{ground-truth} by counting sent packets and compare it with the connection's \emph{potential} sending rate, determined by the receive and network congestion windows.
 
If we could directly monitor the TCP send buffer inside the VM, we could easily observe how much data the application writes to the buffer, how quickly the buffer fills, and the maximum buffer size.  In the absence of OS and application logs from tenant VMs, we rely on two independent metrics to see if the application is sending ``too little" or ``too late'': (i) \emph{generated segment sizes}, to measure if an application is sending too little, and (ii) \emph{application's reaction time}, to see if it is taking too long. 

\begin{figure}
\centering
\subfloat{\includegraphics[width=0.8\columnwidth]{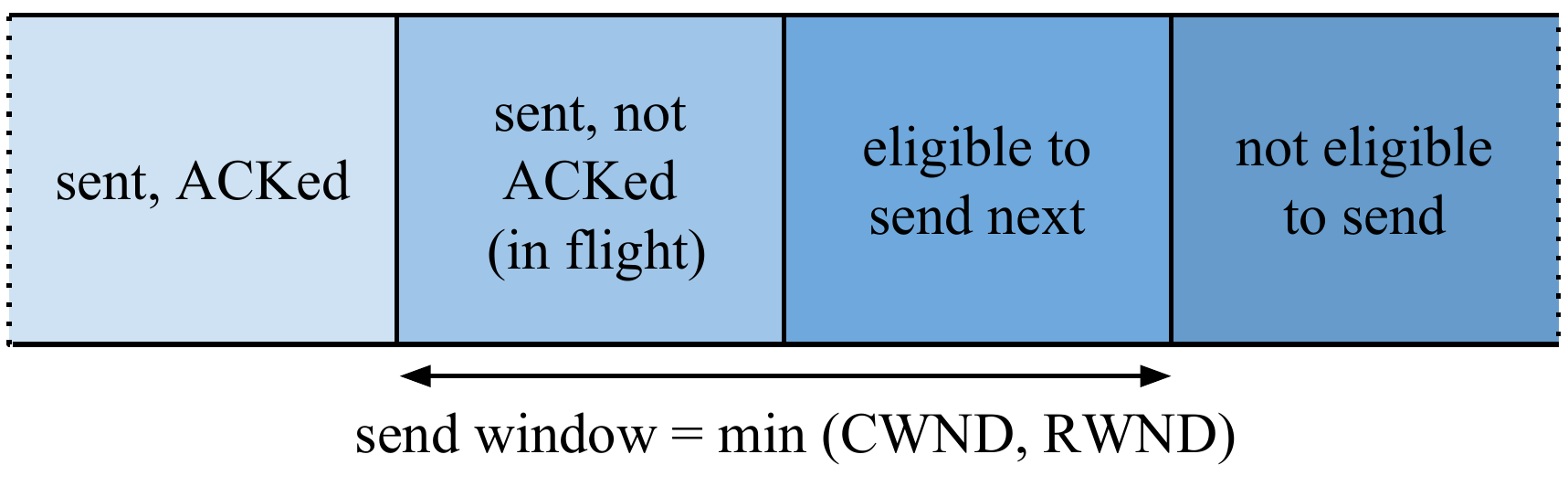}}
\caption{Tracking a TCP connection's flight size}
\label{fig:flight_size}
\end{figure}

\vspace{2mm}
\noindent\textbf{Inferring flight size to measure sending rate:}
Upon transmission of new packets, we update the packet and byte counters for each connection. We also measure the connection's flight size to infer its sending rate (i.e., $\frac{\textit{flight size}}{RTT}$). The flight size of a connection is the number of outstanding packets---packets sent but not ACKed yet---as shown in Figure~\ref{fig:flight_size}, and is inferred via examining the sequence number of outgoing packets and the incoming acknowledgment numbers to track how many segments are still in flight. In this figure, the ``send window'' represents the available window to the sender, that is, the maximum packets that the sender can send before waiting for a new ACK; this window is limited by both the RWND and CWND. In this example, the application is not backlogged, because it has not fully used the available window. We can see that flight size can at most be equal to send window.

Note that the flight size of a connection is an important metric in diagnosis because it depends on the congestion window, the receive window, and the application's own data generation rate (i.e., is the application backlogged?). We will revisit this metric in Section~\ref{sec:net}. 

\begin{figure}
\centering
\subfloat{\includegraphics[height =2.8in]{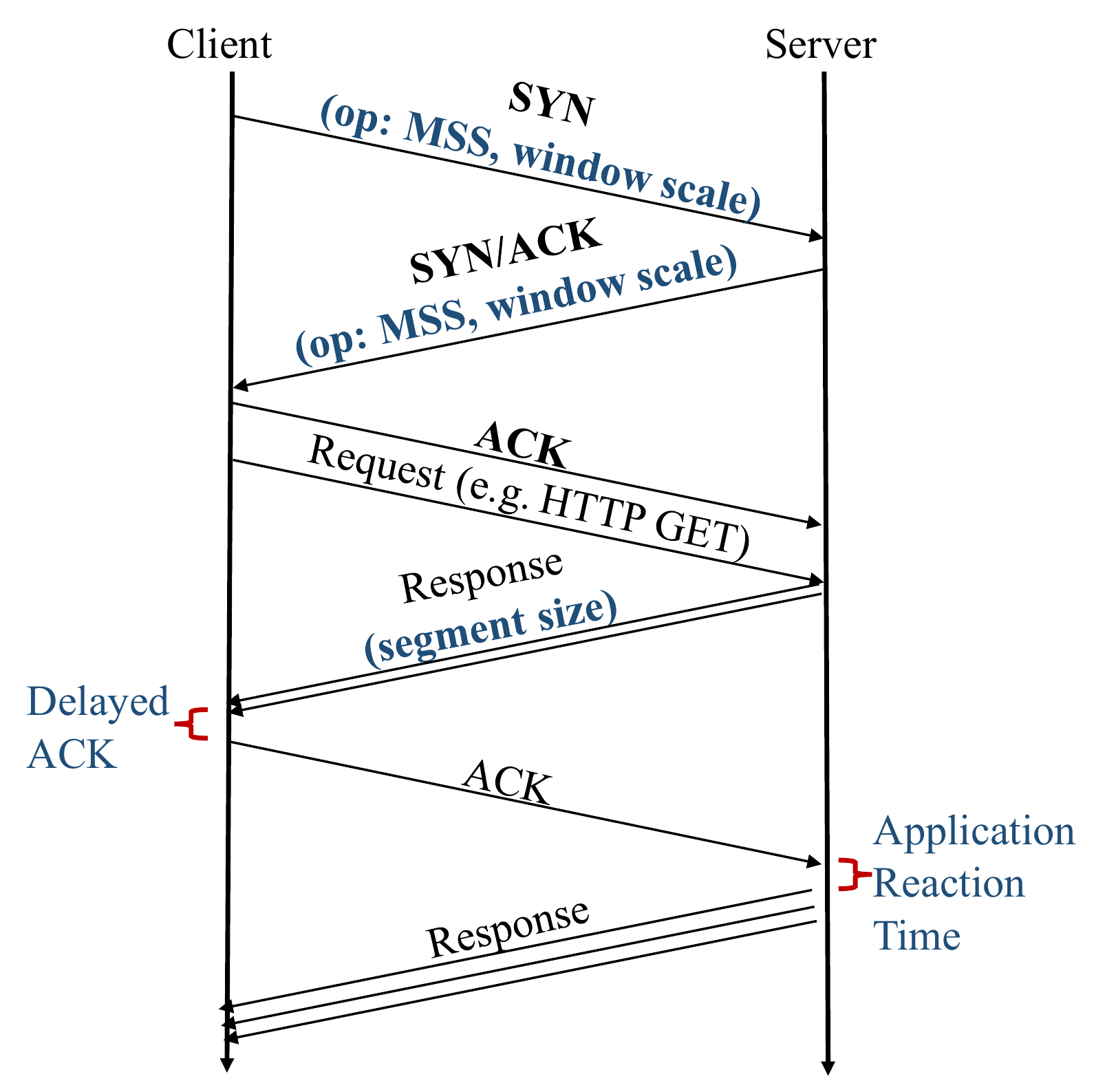}}
\caption{Tracking options, segment size, and application reaction time for a simplex connection (server-client) }
\label{fig:handshake}
\end{figure}

\vspace{2mm}
\noindent\textbf{Extracting MSS from TCP options:} 
The segment sizes in a connection indicate the amount of data that the sending application generates.  We infer MSS by parsing the TCP options exchanged during the three-way handshake, as shown in Figure~\ref{fig:handshake}, in the SYN and SYN-ACK packets.

\vspace{2mm}
\noindent\textbf{Measuring sender's reaction time via cross-packet analysis:} 
We define the sender's reaction time to be the time window between the arrival of a new acknowledgment and the transmission of a new segment of data. The  reaction time evaluates the sending application's data generation rate, i.e., whether or not it is backlogged. Lower reaction times indicate that the data was already processed and was just awaiting an opportunity to be sent. We measure the reaction time by using time-stamps of incoming acknowledgments and outgoing packets. We compare the reaction time with an empirically derived threshold, calculated based on the latency between the edge and the VM. 

Note that ``cross-packet analysis'' happens at the edge, thus the reaction time consists of the application's own data generation latency plus the communication latency between the application and edge (i.e., latency in the hypervisor and virtual switch). More importantly, the network latency does not influence this metric.  Also, notice that the edge acts as \emph{the single point of observation}, seeing both directions of the flow to do cross-packet analysis. These two conditions are not necessarily true on the switches in the core of network to measure the sender's reaction time since the packets could have been delayed at earlier network hops---as opposed to the application itself---or worse yet, taken different paths. 

\subsection{Inferring Network Statistics}
\label{sec:net}
Network problems cause poor performance in a TCP connection. For example, in a congested network, the increased packet loss and path latency cause the TCP congestion control to decrease the sending rate at the sender. The sender's reaction to congestion varies based on the congestion control in use (e.g., Reno vs Cubic) and the severity of the congestion itself.  To determine and quantify the network limitation in the performance of a TCP connection, we need to measure and compare its congestion window---how much the network allows the connection to send---against the receive window---how much the receiver allows the connection to send---and how much data the sender has available to send.

To infer a connection's congestion window, we ``imitate'' the VM's congestion control algorithm by tracking flight size, packet losses and their kinds, and duplicate ACKs. 
We also track RTT and RTTvar to help
pinpoint the effect of network congestion, routing changes, and queuing delay on TCP performance. 

\vspace{2mm}
\noindent\textbf{Inferring loss via retransmission:} We calculate a connection's loss rate  by counting the lost and total packets in the flow.  In addition, we use a counter to track the duplicate ACKs and use it to infer the kind of loss: fast-retransmission (FR) is triggered after a fixed number of duplicate ACKs are received (normally 3); however, a timeout is triggered when no packet has arrived for a while (RTO).  
We track the sequence number of outgoing packets (i.e., packets sent by the server) which helps us find the retransmission of previously seen sequence numbers.  Using the duplicate ACK counter, if we see at least three duplicate acknowledgments before a re-transmission, we treat it as a fast retransmission, otherwise, we deduce that the loss was recovered by a timeout.

\vspace{2mm}
\noindent\textbf{Estimating latency by passive RTT measurements:} We use Karn's algorithm~\cite{Karn:1987:IRT:55482.55484} to estimate SRTT, SRTTvar, and RTO using a series of ``RTT measurements''. An RTT measurement is the time between when a segment was sent and when its acknowledgment reached the sender.  For any connection, upon transmission of a new segment, we create a (time-stamp, sequence number) tuple and maintain it in a queue. Upon arrival of an ACK, we inspect the queue to see if any of the tuples are acknowledged by it. If so, we create an RTT measurement and use it to update the latency statistics via Karn's algorithm.

Note that a retransmitted packet cannot be used as an RTT measurement, because the corresponding ACK cannot be correctly mapped to a single outgoing time-stamp. Also, if a connection has multiple outstanding packets, the queue will have multiple tuples, i.e., the length of the queue grows with the flight size. Finally, if an incoming ACK acknowledges multiple tuples (i.e., a delayed ACK) we de-queue multiple tuples but only create one RTT measurement, from the most recent tuple, to exclude the effect of delayed ACK on RTT. 

\vspace{2mm}
\noindent\textbf{Estimating congestion window via flight size and loss:} 
It is challenging to estimate a connection's congestion window outside the VM's networking stack due to following reasons: 
\emph{(1) Many TCP variations:}
There are many different TCP congestion control algorithms used today (e.g., Reno, New Reno, Cubic, DCTCP). Some are combined with tuning algorithms (e.g., Cubic combined with HyStart to tune \texttt{ssthresh}) or have configurable parameters (e.g., initial window). \emph{(2) Thresholds change:}  \texttt{ssthresh} is the threshold that separates slow-start (SS) from congestion-avoidance (CA) in the TCP state machine and is initially set to a predefined value. However, the Linux kernel \emph{caches} the \texttt{ssthresh} value to use it for similar connections in the future. Also, \texttt{ssthresh} changes throughout the life of a connection (e.g., under packet loss). Therefore, if the full history of a connection (and even \emph{past} connections!) is not observed, \texttt{ssthresh} is unknown, making it impossible to detect transitions from SS to CA based on the \texttt{ssthresh} threshold. 

In the presence of these challenges, we rely on these TCP invariants to infer congestion window: 
\emph{(1)}  The \emph{flight size} of a connection is bounded by congestion window, as shown in Figure~\ref{fig:flight_size}. We denote this lower-bound estimate of congestion window with \cwnd{}. In a loss-free network, \cwnd{} is a moving maximum of flight size of the connection. Note that in the absence of loss, if the connection's flight size decreases, it is either due to the sender producing less data (i.e., not fully utilizing the window) or the receiver's limited receive window; hence, \cwnd{} does not decrease. \emph{(2)} If a packet is \emph{lost}, we adjust \cwnd{} based on the nature of loss, a timeout resets it to IW and a fast-retransmit causes a multiplicative decrease. 

Estimating congestion window based on these invariants makes \cwnd{} ``self-adjustable'', that works regardless of TCP variant and configuration, and is calculated according to Algorithm~\ref{algo:mincwnd}.
Algorithm~\ref{algo:mincwnd} consists of a while loop that inspects every new packet. If a new segment is transmitted and the connection's flight size grows beyond \cwnd{}, we update \cwnd{} to hold the new maximum value of flight size. In case of retransmissions (loss), we decrease \cwnd{} by the multiplicative decrease constant, $C$, if loss is recovered by fast recovery. Otherwise, \cwnd{} is reset to initial window if recovered by a timeout. In this algorithm, we assume the cloud provider knows the value of $C$ and $IW$, since they can be easily inferred, either indirectly via observing how large the first window is and how it changes after a loss, or directly via tools such as Nmap~\cite{Nmap}.

Note that \cwnd{} as estimated by Algorithm~\ref{algo:mincwnd} does not require full knowledge of connection's history, thresholds, or the congestion control algorithm, and is only dependent upon measuring the connection's flight size and loss, thus solves the challenges above without tenant's cooperation\footnote{To keep our heuristics general across all TCP variants, we do not rely on selective acknowledgments.}.
\begin{algorithm}[t]
\DontPrintSemicolon
 \KwIn{multiplicative decrease factor (C), initial window (IW)}
 \KwOut{\cwnd{}}
 \While{ P $\longleftarrow$ \textit{capture new packet} }{
  \If{P is new segment \textbf{and} flight size > \cwnd{}}{
	\cwnd{}$\longleftarrow \textit{flight size}$\;
   }
  \ElseIf{P is retransmitted}
  { 
	  \If{fast retransmit}
	  {
	    \If{first loss in fast recovery}
	   { \cwnd{}$\longleftarrow C\times$\cwnd{}\;}
	  }
	  \ElseIf{timeout}
	  {
	  \cwnd{}$\longleftarrow IW$
	  }
  }
 }
 \caption{Estimating \cwnd{}}
 \label{algo:mincwnd}
\end{algorithm}

\subsection{Inferring Receiver Statistics} 
The receiver-side of a TCP connection can limit the flow by decreasing its advertised window (i.e., RWND) or slowing down the rate of acknowledgments~\cite{SNAP} to control the release of new segments.  

\vspace{2mm}
\noindent\textbf{Tracking RWND per-packet and per-connection :} 
To quantify the receiver limitation in a TCP connection, we track the advertised RWND value \emph{per-packet}, reflecting how much buffer is available on the client. We also track the \emph{per-connection} agreed upon window scaling option during the TCP handshake, as shown in Figure~\ref{fig:handshake}, which is used for scaling RWND.

\vspace{2mm}
\noindent\textbf{Inferring delayed ACKs via RTT samples:}
When an incoming ACK acknowledges multiple tuples in the queue, the ACK must be a delayed ACK, as the client is acknowledging multiple segments at once. When we de-queue tuple(s) based on an incoming ACK, we count and average the number of de-queued tuples per ACK to reflect the effect of delayed ACK.

\begin{figure}
\centering
\subfloat{\includegraphics[width=0.8\columnwidth]{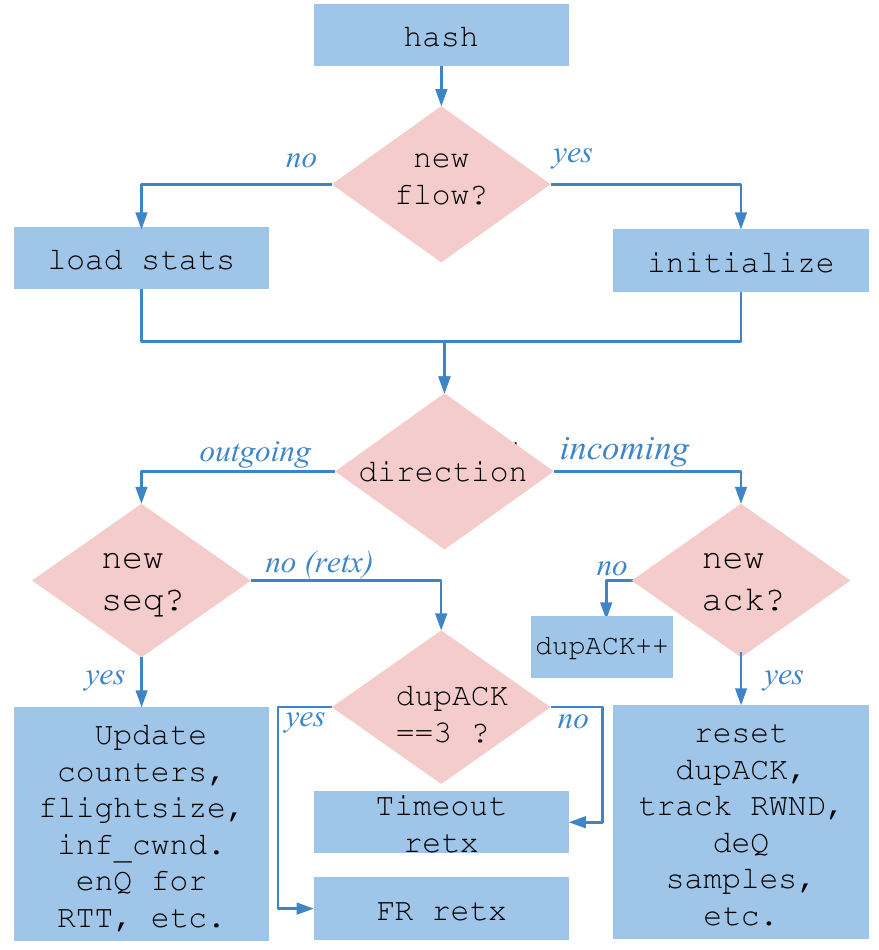}}
\caption{\XYZ's packet-processing logic}
\label{fig:flowchart}
\end{figure}

\vspace{2mm}
\noindent\textbf{Summary:} Figure~\ref{fig:flowchart} summarizes how \XYZ updates a TCP connection's performance statistics while processing a new packet. The packets are first hashed on the four-tuple to either initialize a new flow, or read the existing statistics. Then, based on the direction of the packet, the relevant header fields are extracted and used to update the metrics. The blue boxes show the analytics performed to keep per-flow state (e.g., update flight size) and the pink diamonds show the conditions used to decide which state to update. 

\section{TCP Diagnosis Techniques}
\label{s:diagnosis}
In this section, we describe how \XYZ uses the statistics gathered by the streaming algorithms discussed in section~\ref{s:infer}. 
The high-level objective of the diagnosis techniques is to troubleshoot a connection's performance limitation.  

\vspace{3mm}
\noindent\textbf{Diagnosing Sender Problems:}
Our goal is to find if the sender-side is not limiting the connection's performance (backlogged), or limiting the sending rate via not having enough data to send or taking too long to produce it (non-backlogged).  

\vspace{1mm}
\noindent\emph{1. Exponential sending rate indicates a backlogged sender:} On transmission of new segments, we first examine the connection's macroscopic behavior, i.e., the sending rate, and check to see if it is growing exponentially to infer if the connection is in slow start.  More accurately, we compare the relationship between ACKs and the data packets to see how many packets the sender transmits after a new ACK. If sending rate grows exponentially, we know the connection is \emph{not sender-limited}. Otherwise, we attempt to understand if the connection is sender-limited by checking the next heuristics.  

\vspace{1mm}
\noindent\emph{2. If the sender is backlogged, it will ``completely'' use the send window:} When sending rate does not grow exponentially, the connection could either be in congestion avoidance with a backlogged sender, or it could suffer from a non-backlogged sender, producing less data than CWND.  This heuristic checks to see if the connection's flight size is consistently less than the allowed window to send, determined by the minimum of RWND and CWND, i.e., if $flight size < min(inf\_cwnd, RWND)$. If so, the connection's performance is limited because the sending application does not send more, not because it's not allowed to. 

\vspace{1mm}
\noindent\emph{3. Sending less than allowed, or later than allowed, indicates a non-backlogged sender:} Here we examine the connection's microscopic behavior to see if the connection is under-utilizing the network, not concerning the ``number'' of packets in flight like the previous heuristic, but instead focusing on the ``size'' and ``timing'' of packets.  More concretely, we check to see if a connection is sending packets that are smaller than MSS, or if the application's reaction time (i.e., data generation time) is larger than an empirically derived threshold for backlogged applications. If either of these conditions are met, we conclude that the application is non-backlogged, hence the connection is \emph{sender-limited} indicating that the sender is either not generating enough data to fill up a whole packet, or not responding almost immediately when it is allowed to send.

\vspace{1mm}

\noindent\emph{4. How the flight size changes during a loss recovery gives us a clue of how backlogged the sender is:} In addition to the heuristics above, during loss recovery a connection reveals some information about its internal state\footnote{This heuristic can be treated as a bonus, and the diagnosis algorithm does not rely on seeing a loss.}. As a reminder, fast-recovery causes the congestion window to decrease by a multiplicative factor, $C$. 

\begin{figure}
\centering
\subfloat{\includegraphics[height=2in]{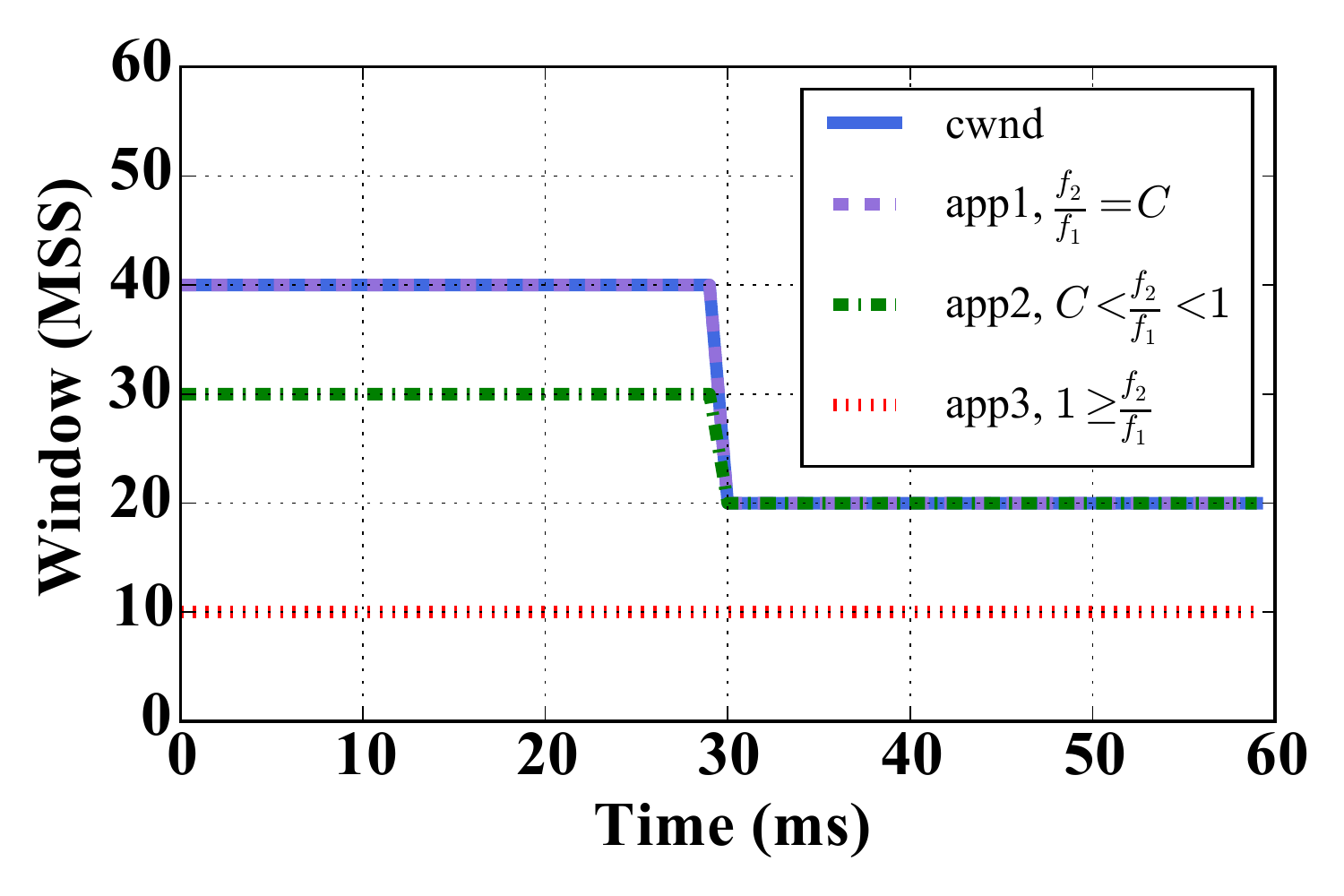}}
\caption{Flight size before and after loss}
\label{fig:loss_flightsize}
\end{figure}

Consider a connection not limited by receiver, and assume that the sending application's data generation rate remains unchanged during the network loss.
We denote the flight size of the connection before loss by $f_1$ and after the loss by $f_2$. Figure~\ref{fig:loss_flightsize} shows three example scenarios, where loss happens at 30ms, prompting CNWD to decrease by half. The $\frac{f_2}{f_1}$ ratio gives us the following insights: if flight size is closely tracking CWND,  $\frac{f_2}{f_1} = C$, the sender is backlogged (app 1); if the connection was not fully using the CWND before loss but is backlogged after the loss, $C < \frac{f_2}{f_1} < 1$ (app 2 ); finally, if the flight size remains unchanged the sender is not backlogged (app 3). Note that in these examples we assume the state of the sender remains unchanged during the loss recovery.

\vspace{3mm}
\noindent\textbf{Diagnosing Network Problems:}
Our goal is to determine if the network is restricting TCP performance, either due to limited bandwidth (congestion window limited), high packet loss rate, or increased latency due to problems such as queuing delay or routing problems.

\vspace{1mm}
\noindent\emph{1.  Small congestion window hurts TCP's performance:}
When the network has performance constraints, for example limited network bandwidth, the congestion window will limit the rate of the connection, that is: $\textit{flight size} \leq inf\_cwnd < RWND$.  Upon a packet retransmission, if loss causes the \cwnd{} value to drop below the RWND,  we deduce the connection is limited by network. 

\vspace{1mm}
\noindent\emph{2.  Increased network path latency slows TCP's rate:}
The sending rate of the connection is a function of both the flight size and RTT; the sender can only increase the window after a new ACKs arrives, which usually takes an RTT. 
To track the impact of network latency on TCP performance, we use the RTT measurements as explained in section~\ref{s:infer}.  The cloud provider can either define an ``expected RTT'' per TCP connection based on SLAs, or use the minimum RTT sample per-flow as the baseline. To diagnose path latency problems, we compare the RTT values with the expected RTT to  detect if a connection's latency is acceptable. 

\vspace{3mm}
\noindent\textbf{Diagnosing Receiver Problems:} 
Our goal is to find if the receiver is restricting TCP performance, either by offering a small receive buffer (receive window), or by delaying ACKs.

\vspace{1mm}
\noindent\emph{1. Small receive window hurts performance:}
Upon updating the RWND and \cwnd{} values, this heuristic compares them with the current flight size to see if the connection's sending rate is receiver-limited, that is: $\textit{flight size} \leq RWND < inf\_cwnd$. If so, the connection is diagnosed as receiver-limited. 

\vspace{1mm}
\noindent\emph{2. Delayed acknowledgment hurts performance:}
The receiver can limit TCP performance by sending ACKs with a delay; delayed ACKs has been shown to cause issues in datacenters~\cite{SNAP}. When the client sends ACKs with a delay, for example, sending acknowledgments for every other packet, the sender's opportunity to increase its window is halved\footnote{Note that the congestion window on sender-side opens up upon receiving each new ACK, as every ACK is a sign that a packet has left the network, hence the network can receive more.}. For each connection, we measure the average number of RTT samples freed by each new ACK and if the average is greater than one, we diagnose the connection as receiver-limited due to delayed acknowledgment.

\section{Data-Plane Monitoring}
\label{s:hardware}
In this section, we describe how \XYZ tracks TCP connections in the data plane and discuss the principles behind our target-independent solution using P4. We outline the P4 features (e.g., metadata) that enable us to monitor TCP connections according to Figure~\ref{fig:flowchart} (Section~\ref{s:prototype}); then, we discuss the target-specific resource constraints and how to mitigate them (Section~\ref{s:limit}).


\subsection{TCP Monitoring Prototype in P4}
\label{s:prototype}
P4 is a programming language that allows us to express how packets are processed and forwarded in a target-independent program, therefore, our P4 prototype can run in a public cloud on a variety of targets, as long as at least one of the elements at the edge (the switch, the hypervisor, or the NIC) can run P4 programs.

To monitor TCP connections in the data plane in real time, we need to extract and retain packet header information (P4's flexible parsing), carry information across multiple stages of packet processing (P4's metdata), and store state across successive packets of the flow (P4's registers).
Furthermore, to realize the logic in Figure~\ref{fig:flowchart}, we need to perform specific operations on each packet, shown with blue boxes (P4's tables and actions) and check test conditions based on both the packet headers and the flow state to invoke the relevant tables, shown by pink diamonds (P4's flow control). 

\vspace{2mm}
\noindent\textbf{1.	Extracting headers and options via flexible parsing:} 
Using header definitions, we identify the relevant header fields in a packet. In our prototype we assume the TCP packets have Ethernet, IPv4, and TCP headers, although this can be easily extended to include other protocols (e.g., IPv6). The following snippet shows some relevant TCP headers. In addition, we provide a parser that extracts headers (e.g., source and destination IP from the IP header).
\begin{lstlisting}[xleftmargin=.1\textwidth, xrightmargin=.1\textwidth]
header_type tcp_t {
    fields {
        srcPort : 16;
        dstPort : 16;
        seqNo : 32;
        ackNo : 32;
        ...
        }
}
\end{lstlisting}

P4 models the parser as a state machine represented by a parse graph. The parsed headers need to be ``de-parsed'', i.e., merged back, to a serial stream of bytes before forwarding.  TCP options require TLV (Type-Length-Value) parsing. For parsing options, we use ``masks'' to identify the ``type'' (e.g., type 2 represent MSS), then a parser is called to extract that option knowing its ``length'' (e.g., \texttt{parse\_mss} for MSS in the snippet below) which returns the control back to the original parser when done. This creates a loop in the parsing graph causing the exact de-parsing behavior to be undefined. To solve it, we impose a fixed order for de-parsing, by using the \textit{pragma} keyword as the following code snippet shows. 
We will use this extracted MSS value in the subsequent parts.  

\begin{lstlisting} 
@pragma header_ordering ethernet ipv4 tcp options_mss options_sack options_ts options_nop options_wscale options_end

parser parse_tcp_options {
	return select(mymeta.opt_counter, current(0,8)) {
    	...
		0x0002 mask 0x00ff : parse_mss;	
        ...
	}
}
\end{lstlisting}
\vspace{2mm}
\noindent\textbf{2.	Keeping per-flow state in registers:}
Registers are stateful memories, which are essential to \XYZ because they maintain the per-flow state as it gets updated after processing each packet. Registers consume resources on the target, hence are a major limitation in running our solution on specific targets, therefore we will minimize the required per-flow state to ensure our program runs on commodity hardware in Section~\ref{s:mid}.

P4 registers can be global, referenced by any table, or static, bound to a single table. The following code shows one of our global registers, \texttt{MSS}, as an array of 16-bit values, and \texttt{instance\_count} is the number of entries in the flow table. Each packet is hashed to find its flow index in the register array. We will explain our bi-directional flow hashing in more details shortly.
\begin{lstlisting}[xleftmargin=.1\textwidth, xrightmargin=.1\textwidth]
register MSS {
    width : 16;
    instance_count : ENTRIES;
}
\end{lstlisting} 
When tracking a TCP connection for diagnosis, some register values depend on the value of other registers; for example, only by comparing a packet's acknowledgment with previous ACKs of the flow can we detect a duplicate ACK. To update such dependent registers, we have to \emph{read} other register(s), test \emph{conditions}, and finally \emph{update} the target register.  

\vspace{2mm}
\noindent\textbf{3.	Carrying information per-packet via metadata:}
Metadata is the state associated with each packet, not necessarily derived from the packet headers, and can be used as temporary variables in the program. We use metadata to carry the information belonging to the same packet from one table to the other. The code below shows the most widely used metadata in our program, \texttt{flow\_map\_index}, which is the flow's index produced by hashing. This metadata carries the index over to the subsequent tables, each using it to index their registers for read/write.  Below, we show a code snippet for declaring metadata field named \texttt{flow\_map\_index} where \texttt{FLOW\_MAP\_SIZE} indicates its width in bits. In the next subsection, we explain how we use this metadata in hashing.

\begin{lstlisting}
header_type stats_metadata_t {
	fields {
    flow_map_index : FLOW_MAP_SIZE; // flow's map index
    ...
   }
}
metadata stats_metadata_t stats_metadata;
\end{lstlisting} 

Some metadata has special significance to the operation of the target (i.e., the standard intrinsic metadata).  In particular, we use the target's \texttt{ingress\_global\_timestamp} as the arrival time of the packet, which is necessary to infer latency metrics such as the sender's reaction time and SRTT.

\vspace{2mm}
\noindent\textbf{4.	Bi-directional hashing using metadata and registers:}
As discussed earlier, our streaming algorithm must see both directions of traffic to capture our cross-packet metrics, e.g., application reaction time.
To do this, we need to hash both directions to the same index and process them as one entity. Unfortunately, P4 provides no primitives or methods for hashing both directions to the same index---no symmetric hashes. Although some targets may allow configuring the hash function through run-time APIs, this support may vary across targets~\cite{p4spec}. Therefore, we build our own symmetric hash using P4's default hash algorithm, e.g., \texttt{crc32}, by defining two sets of headers to hash on, with one in the reverse order of the other. In other words, one direction is hashed based on (src IP, dst IP, src Port, dst Port) fields, and the reverse direction is hashed on (dst IP, src IP, dst Port, src Port) fields. To keep the direction's hash function consistent, we use a simple and consistent comparison of the two IPs: if $src IP > dst IP$, we hash the packet header in the former order, otherwise we hash the packet headers in latter order. This guarantees that each side of packet stream gets consistently hashed by one of these hash functions, but results in the same index value per flow. 

\vspace{2mm}
\noindent\textbf{5. Realizing operations using actions and tables:}
To realize the blue boxes in the flowchart of Figure~\ref{fig:flowchart}, P4 tables and actions are used. 
A P4 table defines the fields to match on and the action(s) to take, should the entry match. 
P4 tables allow us to express different sets of match-action rules to apply on packets; for example, the set of actions for an outgoing packet differs from incoming packets. Furthermore, some tables could be dedicated to monitoring while others are dedicated to forwarding packets (e.g., \texttt{ipv4\_lpm} and \texttt{forward}). A fundamental difference between our monitoring tables from regular forwarding tables in P4 is that our monitoring tables have a single static entry that matches on every packet ---hence, have no match field. In contrast, \texttt{ipv4\_lpm} is a forwarding table that uses longest prefix matching to find the next hop.  The following code snippet shows two of \XYZ's tables, the \texttt{lookup} table, that hashes every packet to find its flow index, and the \texttt{init} table that initializes the flow upon observing its first packet (e.g., saves the extracted MSS value in the MSS register array, at the flow's index).  

\begin{lstlisting}[xleftmargin=.1\textwidth, xrightmargin=.1\textwidth]
table lookup{
	actions {
		lookup_flow_map;
	}
}
table init{
	actions {
		init_actions;
	}
}
\end{lstlisting}


\begin{lstlisting}
action lookup_flow_map() {
 modify_field_with_hash_based_offset(stats_metadata.
 flow_map_index, 0, flow_map_hash, FLOW_MAP_SIZE);
}

action init_actions() {
	register_write(MSS, stats_metadata.flow_map_index, options_mss.MSS);
    ...
}

\end{lstlisting}

Actions in P4 are declared imperatively as functions, inside the tables. 
Actions can use registers, headers, and metadata to compute values. An example action is \texttt{register\_write}, which takes a register, and index, and a value as input, and sets the value of the register array at the index accordingly. Actions are shown with red color in our code snippets.

\vspace{2mm}
\noindent\textbf{6. Conditions via control-flow:}
The control flow of a P4 program specifies in what order the tables are to be applied. Inside the control segment, we can ``apply'' tables and test conditions. 
The choice of which block to execute may be determined by the actions performed on the packet in earlier stages. The control flow is what enables us to design the pipeline and implement the conditions (pink diamonds) in P4 as the flowchart shows in Figure~\ref{fig:flowchart}. 

In the ``widely-supported'' P4 specification~\cite{p4spec}, conditional operations are restricted to the control segments of program; that is, we cannot have \texttt{if-else} statements inside a table's logic. Fortunately, P4 offers metadata, which can be used as temporary variables in the program. The metadata gives us an opportunity to read the current value of \emph{conditional registers} inside an earlier table---the ``loader''---in the pipeline, store their values in the metadata, test the conditions in the control section, and apply the appropriate set of tables conditionally. 
Note that we need the ``loader'' table because of the current restrictions in P4 that allows conditions only in the control segment. 

\begin{lstlisting}
control ingress {
	if ( ipv4.protocol == TCP_PROTO) {
		if( ipv4.srcAddr > ipv4.dstAddr ) {
			apply(lookup);
		}else{
			apply(lookup_reverse);
		}	
		if ( (tcp.syn == 1) and (tcp.ack == 0) )//first pkt
			apply(init);
		else
			apply(loader);
		if (ipv4.srcAddr == stats_metadata.senderIP){	
			if( tcp.seqNo > stats_metadata.seqNo ){
				apply(flow_sent);
				if(stats_metadata.sample_rtt_seq == 0)
					apply(sample_rtt_sent);//"temp" has the new flightsize
				if(stats_metadata.temp > stats_metadata.mincwnd)
					apply(increase_cwnd);					
			}else{
				if(stats_metadata.dupack == DUP_ACK_CNT_RETX)
					apply(flow_retx_3dupack);
				else
					apply(flow_retx_timeout);
			}
		}
		else if(ipv4.dstAddr == stats_metadata.senderIP ) {
			if( tcp.ackNo > stats_metadata.ackNo ){
				apply(flow_rcvd);//new ack
				if( tcp.ackNo >= stats_metadata.sample_rtt_seq and stats_metadata.sample_rtt_seq>0){
					if(stats_metadata.rtt_samples ==0)
						apply(first_rtt_sample);
					else
						apply(sample_rtt_rcvd);
				}
			}else
				apply(flow_dupack);//duplicate ack
		}
	}
    apply(ipv4_lpm);
    apply(forward);
}
\end{lstlisting}

\subsection{Hardware Resource Constraints}
\label{s:limit}
In this section, we explain our design choices to monitor connections in P4. These choices stem from a variety of restrictions, in particular, the limited resources on hardware switches, missing features in the P4 spec, and the diversity of hardware targets, which would require us to design for the least common denominator among the supported features. 

\vspace{3mm}
\noindent\textbf{1. Handling hash collisions:}
We use 32 bits for hashing in our prototype; regardless, collisions are often a concern in hash tables. In our software implementation, we handle collisions in the hash table by ``hash-chaining'': we store the four tuple key of the connection and create a linked-list of flows in the same index with different keys. 
  However, since memory in hardware is limited, we decide to not store a connection's tuple. 
Still, if collisions go undetected they may pollute the accuracy of collected statistics. Hence, it is useful to assess if the accuracy of a connection's statistics has been compromised. Therefore, we perform basic checks on the packet's sequence number versus the flow's previously sent sequence numbers and available windows (i.e., does the sequence number fall within the acceptable window?). This comparison requires additional tables or registers per-flow, but can store the result of conditions in a Boolean variable, named ``sanity check''. The sanity check can be queried from the data plane along with the connection metrics to indicate whether collected statistics are reliable for diagnosis. 

\vspace{3mm}
\noindent\textbf{2. Keeping one RTT sample at a time:}
As discussed in Section~\ref{s:diagnosis}, to accurately track the connection's RTT, we maintain a queue of tuples based on the outgoing packets, where each tuple is (sequence number, time-stamp). The received ACKs are compared to the tuples of the queue to make an RTT measurement. Since the queue of tuples grows with the flow's flight size, it increases the amount of state per-flow in our P4 program. Because the hardware resources on a switch are limited, we limit the number of outstanding RTT tuples in our P4 program to one at a time, per-flow: we only sample an outgoing packet for RTT if the flow's queue is empty. 

\vspace{3mm}
\noindent\textbf{3. Multiple accesses per register array:}
Our program accesses some registers from multiple points to use them in test conditions; e.g. duplicate ACK count register is read in the loader table and used for identifying the kind of loss, and upon a new duplicate ACK another table updates it. Currently, conditional operations in P4 are restricted to the control segments. Thus we cannot avoid accessing some registers in multiple tables.
Unfortunately, accessing a register from multiple tables limits the processing rate. However, the ternary operator (\texttt{?:}) will be supported in the next P4 version~\cite{chang}, allowing us to perform simple conditional assignments, eliminating the need for global registers, and permitting our solution to run at line rate. 

\vspace{3mm}
\noindent\textbf{4. Relying on control plane to scale RWND:}  
In TCP, the advertised RWND should be shifted by the window scale (as negotiated in handshake) to calculate the actual receive window. However, most P4 targets can only ``shift'' by a ``constant'' value~\cite{p4spec}. So, we instead record both values and allow the control plane to query both and perform the shift.  

 \vspace{3mm}
\noindent\textbf{5. Foregoing RTTvar:} Calculating RTTvar involves capturing the absolute difference of the smoothed moving average RTT (SRTT) and the current RTT sample (RTT). This difference can be captured via an \texttt{abs} operator or by introducing a new comparison test, i.e., a new pipeline stage. Unfortunately, the P4 specification does not support the \texttt{abs} operator and adding a new stage impacts the processing rate of our implementation to gain a single metric.  By default, \XYZ does not include this stage but can be enabled optionally.
\section{Two-Phase TCP Monitoring}
\label{s:mid}
Our goal in this section is to lower the cost of monitoring and diagnosing  TCP connections. We present a two-phase monitoring technique to decrease the amount of state required.
The first phase monitors all connections continuously but only collects low-overhead metrics, enough to \emph{detect} but not \emph{diagnose} performance problems. When a connection meets the ``badness'' criterion, heavier-weight monitoring is enabled to diagnose the poor performance. 

\vspace{3mm}
\noindent\textbf{Phase 1: Lightweight detection:}
In the first phase, we collect lightweight metrics that suffice to ``detect'' the existence of performance problems, based on a badness criterion.  The statistics used in the first phase must be: 1. \emph{lightweight} to maintain, ensuring that the continuous monitoring of connections in the data plane is cheap, and 2. \emph{general} enough to capture the badness of the flow, regardless of the component limiting the performance. We use the \emph{average rate of the flow} as an indicator of how well it's performing. 

To maintain the average rate of flow, we keep three registers: 1. \texttt{init\_time}, the time-stamp when monitoring began, 2. \texttt{bytes\_sent}, total bytes sent so far, and 3. \texttt{update\_time}, the time-stamp when the flow was updated last. The control plane can query these states per-flow and find the connections that look troubled based on a cloud-wide tenant-specific threshold\footnote{If the tenant's traffic is not crossing the WAN.}. Upon observing low rate (or when tenant complains), the cloud provider can turn on the heavy-weight monitoring mode to do diagnosis.

\vspace{3mm}
\noindent\textbf{Phase 2: Diagnosis of troubled connections:}
The second phase is ``diagnosis'', where we collect heavyweight metrics for a troubled connection to shed light on the component that is hindering the flow’s performance. These metrics include the complete set of TCP statistics (discussed in Section~\ref{s:infer}) and our diagnosis techniques (discussed in section~\ref{s:diagnosis}). This phase consumes more state on the switch, but note that it is only turned on after a problem is detected, hence the switch state consumption overall decreases. 
The two-phase monitoring can be thought of as a ``long and narrow'' table (all flows, few metrics), followed by a ``short and fat'' table (few flows, many metrics). 

There are some challenges with inferring TCP metrics midstream: First, the TCP constants, in particular MSS and window scale are exchanged once, during the handshake. Second, the flow counters (e.g., packets sent, ACKed, or in flight) are unknown. This results in errors in the inferred value of flight size, which is also used to estimate \cwnd{}. 

Our solution to these challenges is two fold: 1. Parse and keep the TCP options during the handshake for all the flows in the first phase, in case they are needed later. This approach provides high accuracy but requires more data-plane state. 2. Infer them midstream, only when necessary. Of course, inferring constants midstream reduces the memory overhead at the expense of accuracy.  To infer MSS from midstream, we need to keep track of the largest segment sizes seen so far. To infer the window scaling option midstream we track the flight size and the unscaled RWND as advertised in received packets. We use the TCP invariant that ``the flight size of a connection is limited by the receive window'', hence we can estimate the \emph{lower-bound} window scaling factor:

\begin{gather}
\label{eq:wscale}
\textit{flight size} \leq \textit{RWND} \cdot 2^\textit{scale}\nonumber\\
\ceil {\log _2 \frac{\textit{flightsize}}{\textit{RWND}}} \leq \textit{scale}\label{equation}
\end{gather}

Finally, to infer \cwnd{}, we rely on flight size, which itself is inferred from tracking sequence numbers of outgoing packets and incoming ACKs midstream. We will evaluate the accuracy of these metrics to show how close to actual values the accuracy of midstream inferred metrics get.  
\section{Evaluation}
\label{s:eval}
In this section we evaluate the overhead of our P4 prototype for hardware switches and our C prototype for hypervisors (Section~\ref{s:eval:p4}). Then we use the software prototype to validate the accuracy of our diagnosis algorithm using synthetic traffic (Section~\ref{s:eval:diagnosis}). Next, we showcase \XYZ in the wild by analyzing CAIDA packet traces (Section~\ref{s:eval:caida}). Finally, we demonstrate the trade-offs in accuracy and overhead in the P4 design.

\subsection{CPU and Memory Overhead}
\label{s:eval:p4}
We evaluate \XYZ's overhead along two dimensions: memory utilization on hardware switches, and CPU Utilization on hypervisor or vswitch; this is mainly because switches operate at line rate but are constrained for memory, while software solutions are not often constrained by memory, but by CPU utilization. \XYZ's software prototype is implemented in \texttt{C} and uses \texttt{libpcap} to capture packets. 

\begin{figure}[t!]
\centering
\includegraphics[height=2in]{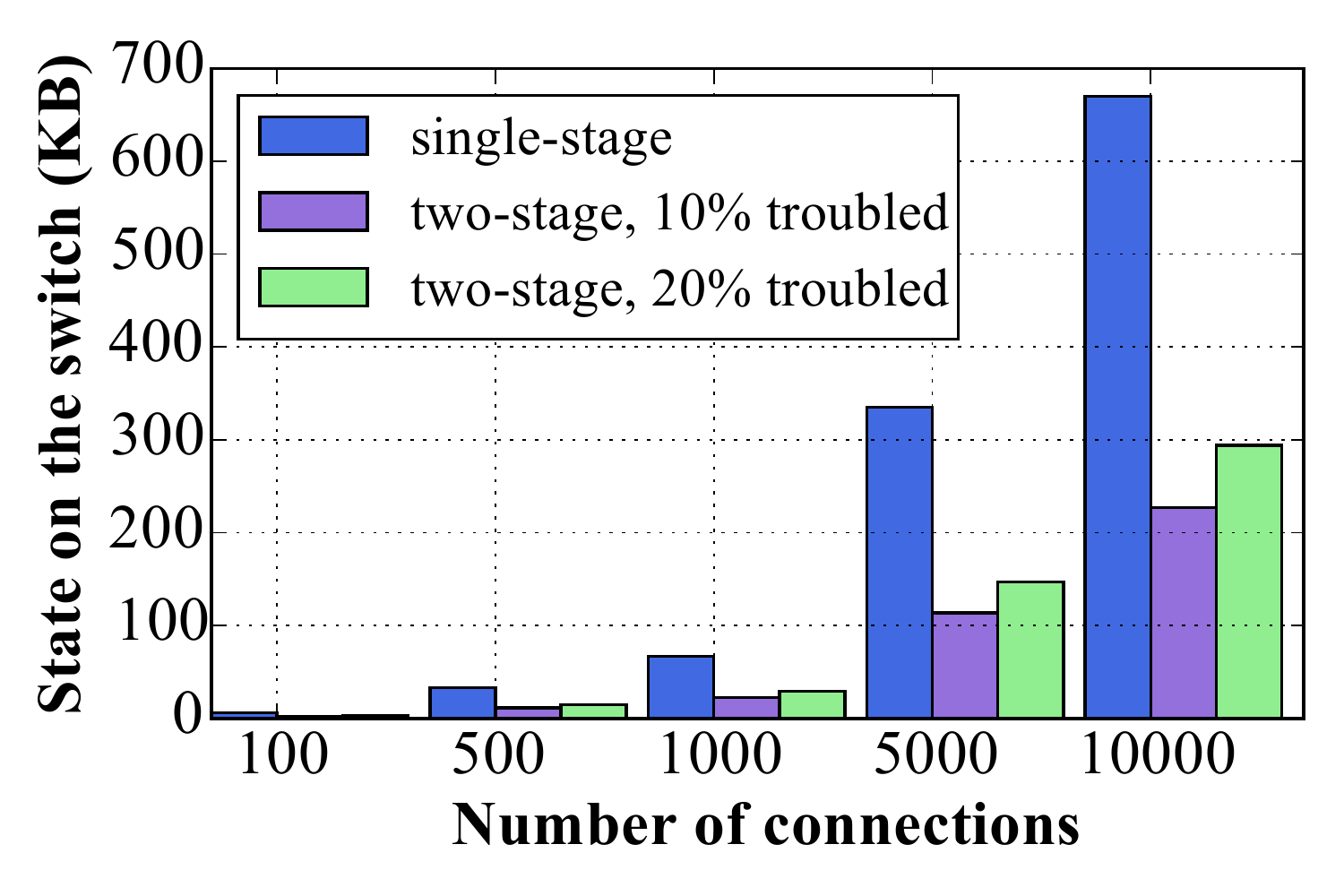}
\caption{Required state on switch in single vs two-phase monitoring, with 10\% and 20\% troubled connections}
\label{fig:memory}
\end{figure}

\vspace{1mm}
\noindent\textbf{Memory in hardware:}  In single-phase mode, our P4 prototype keeps 67 bytes of state for each connection (i.e., 16 four-byte registers to keep the flow state, a two-byte register to track MSS, and a one-byte register for scale). In addition, 40 bytes of metadata are used to carry a packet's information  across tables.  In a typical data center, a host can have 10K connections~\cite{SNAP}, which results in 670 KB of state required to track their state. In the two-phase monitoring prototype, as described in section~\ref{s:mid}, we use the average rate of flows as the badness factor in the first stage, which would impose about 16 bytes per flow in the first stage. Figure~\ref{fig:memory} shows the amount of state needed for single-phase monitoring, as opposed to two-phase monitoring with 10\% and 20\% troubled connections.

\begin{figure}[t!]
\centering
\includegraphics[height=2.1in]{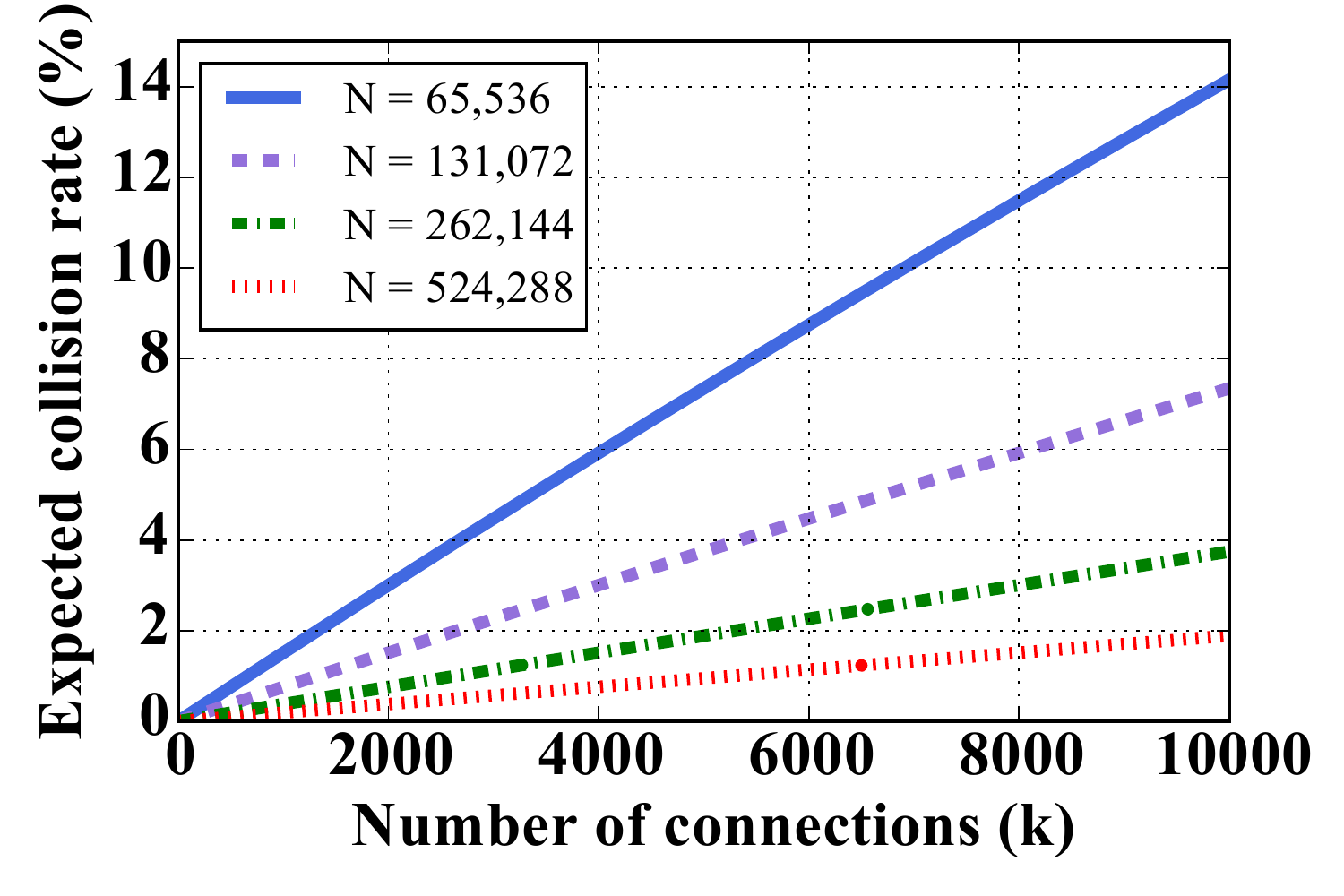}
\caption{Expected collision rate vs number of flows (k) for different table sizes (N).}
\label{fig:collision}
\end{figure}

As more connections are monitored, the expected rate of collision in the hash table increases.  Figure~\ref{fig:collision} shows the expected collision rate per number of flows (k), for varying sizes of tables (N). Assuming the N hash values are equally possible, the probability that a flow A shares the same index with flow B is $\frac{1}{N}$.  So, the probability that the other k-1 flows will not share the same index is $(1-\frac{1}{N})^{k-1}$, resulting in the expected likelihood of collisions of $1-(1-\frac{1}{N})^{k-1}$.
Thus, to track 10K connections with a collision rate of less than 4\%, we need the table size to be at least $262,144$ ($2^{18}$), which results in less than 18 MB space on the switch. 

\begin{figure}[t!]
\centering
\includegraphics[height=2in]{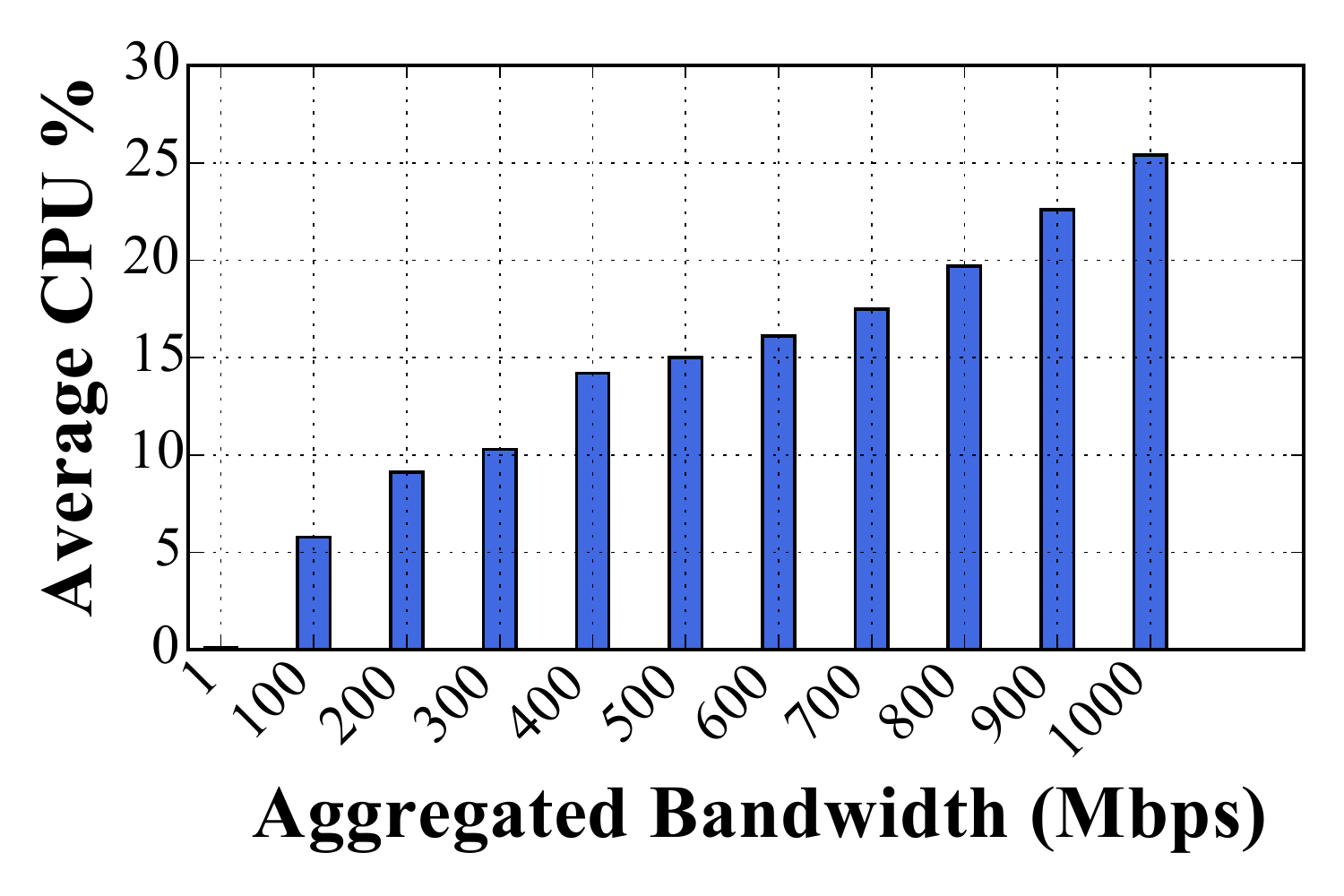}
\caption{CPU per aggregate bandwidth processed}
\label{fig:cpu_overhead}
\end{figure}

\vspace{1mm}
\noindent\textbf{CPU in software:} 
We measure the CPU overhead by connecting two servers to a single switch and starting parallel TCP flows between them and measuring CPU using \texttt{top}. The server machines have Xeon e3-1630 V3, 4 core, 3.4 GHZ processors. All flows are initially established (i.e., completed the TCP handshake) and have the average rate of 1 Mbps. Figure~\ref{fig:cpu_overhead} shows the CPU consumption versus the aggregate bandwidth processed. 
\XYZ's CPU consumption is close to approaches that use near-real time polling frequency (e.g., ~\cite{SNAP} at 50ms frequency). 

\begin{figure}[t!]
\centering
\includegraphics[height=2in]{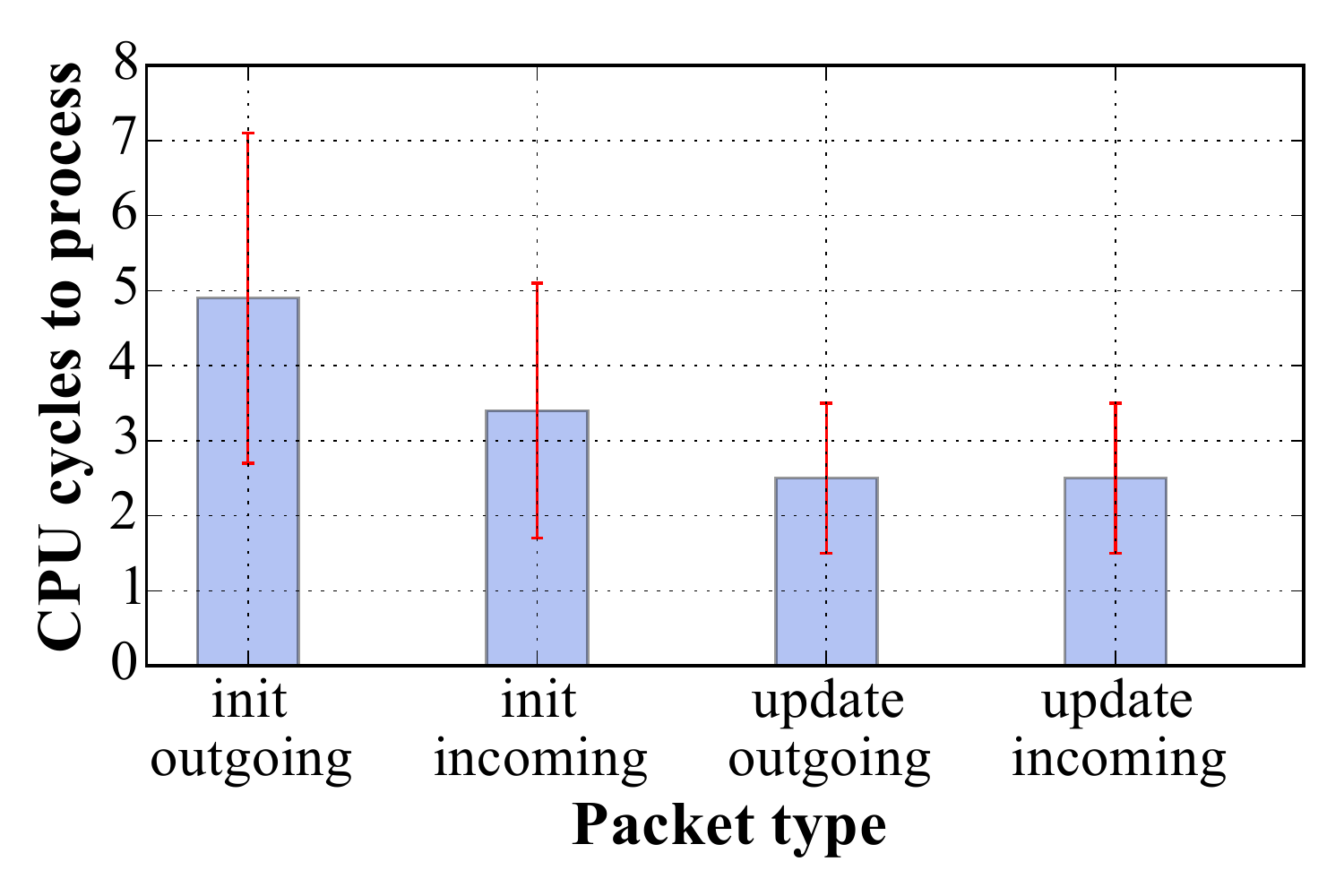}
\caption{CPU cycles to process each type of packet, allocation and initialization impact first packets }
\label{fig:cpu_kind}
\end{figure}

In Figure~\ref{fig:cpu_kind}, we quantify the CPU processing requirements for different packet types. The first packet of a flow usually takes longer to process, because \XYZ must allocate and initialize flow state and parse packet options that require extra CPU cycles. Further, the first outgoing packet is used for creating the first tuple for RTT measurement. The variations in the cycles are caused by several reasons: First, our software prototype's flowtable is a hash table with hash-chaining, thus, in case of a collision, the flow statistics are maintained in a linked-list; incurring the extra overhead of linked-list traversal for collisions. Second, for retransmitted packets or duplicate ACKs, the flow state needs further updates. Finally, packets used for RTT measurement incur extra processing. 

\subsection{Diagnosis Accuracy}
\label{s:eval:diagnosis}
We measure the accuracy of our diagnosis method by systematically creating TCP connections with known problems and comparing the diagnosis results with the ground truth. We create a server-client connection with the client requesting a 1MB file over a link with 1Mbps bandwidth and 50ms round-trip time. We then create the following problems in the connection and repeat the experiments 100 times: 

\noindent\emph{1. Sender-Limited:} We emulate a resource bottleneck for the server, e.g.,  slow disk or busy CPU, by making the server wait for $T$ seconds before transmitting each data packet. Higher $T$ indicates more severe problems. We also emulate non-backlogged servers by limiting the transmitted segment sizes to less than an MSS.

\noindent\emph{2. Receiver-Limited:} We create receiver-limited connections by changing socket options (using Linux's \texttt{setsockopt}) to limit the client's receive buffer size, \texttt{socket.SO\_RCVBUF}.

\noindent\emph{3. Network-Limited:} We use the Gilbert-Elliot model~\cite{5755057} to emulate micro-bursts during network congestion: a connection can be in either a \emph{good} (no network congestion) or \emph{bad} (network congestion) state. To emulate the bad state, we generate bursty losses at a rate of 1\% to 10\% for 2 seconds. We assume that losses in the good state are negligible.

\begin{figure}[t!]
\centering
\includegraphics[height=2in]{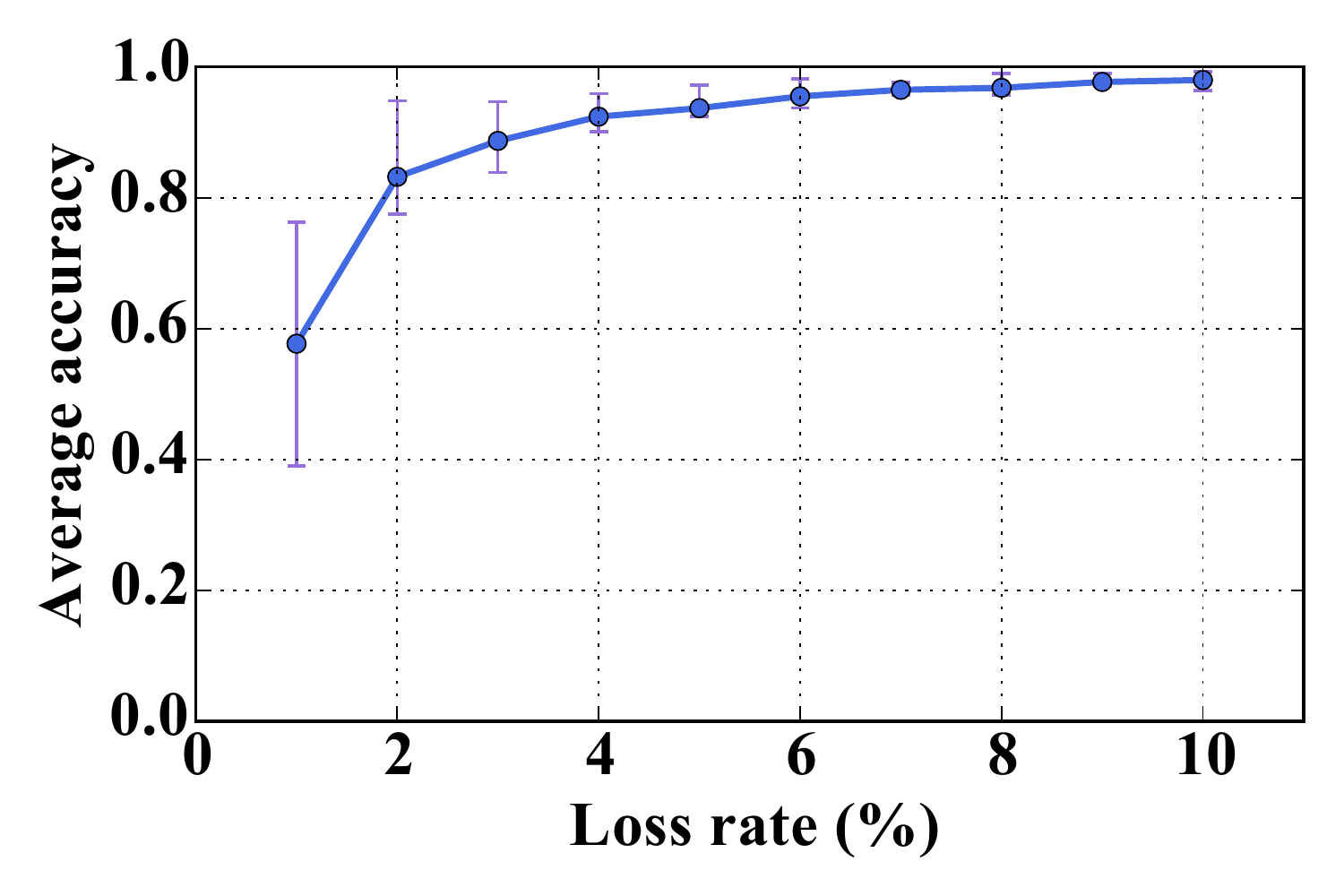}
\caption{Accuracy vs severeness of problem}
\label{fig:accuracy_sever}
\end{figure}

For each problem shown in Table~\ref{table:roc}, we measure ``sensitivity'' and ``accuracy'': the \emph{true positive rate} (TPR) is the fraction of correctly diagnosed tests, showing the sensitivity to each problem, and \emph{diagnosis accuracy} is the fraction of time \XYZ correctly classified the connection to be limited by that problem. The initial results are promising, \XYZ achieves an average accuracy of 94\%; \XYZs accuracy is less than 100\% because the ability to detect a problem is proportional to its severity.  Figure~\ref{fig:accuracy_sever} shows an example problem where the severity changes from low (1\% loss) to high (10\% loss), increasing the average accuracy of diagnosis. 


\begin{table}[t]
\ra{1.3}
\centering
\caption{\XYZs Diagnosis Sensitivity and Accuracy }
\scriptsize
\begin{tabular*}{\columnwidth}{@{}p{4cm}p{2cm}p{3cm}@{}}
\toprule 
\textbf{Problem} & \textbf{TPR} & \textbf{Avg Accuracy} \\
\midrule 

Sender-Limited & $98\%$ & $95\%$ \\

Receiver-Limited & $96\%$& $94\%$ \\

Network-Limited & $94\%$	& $93\%$\\

Sender-Network-Limited & $100\%$ & $95\%$  \\

Receiver-Network-Limited & $100\%$ & $93\%$\\

\bottomrule 
\end{tabular*}
\label{table:roc}
\end{table}
\subsection{Analyzing CAIDA Traces}
\label{s:eval:caida}
In the absence of traces from a public cloud, we use CAIDA traces collected on Equinix-Chicago~\cite{caida} to showcase \XYZ, by assuming that the applications that produced this traffic have now moved to the cloud. We preprocess the trace by eliminating connections with less than 10 packets in either direction. A total of 244,185 flows are considered in this study. We make one
modification in \XYZ to account for the fact that the packet traces are
collected from within the network, rather than the edge:
we turn off inferences based on application reaction time because it cannot be reliably captured in the core of the network.

\begin{figure}[t!]
\centering
\includegraphics[height=2in]{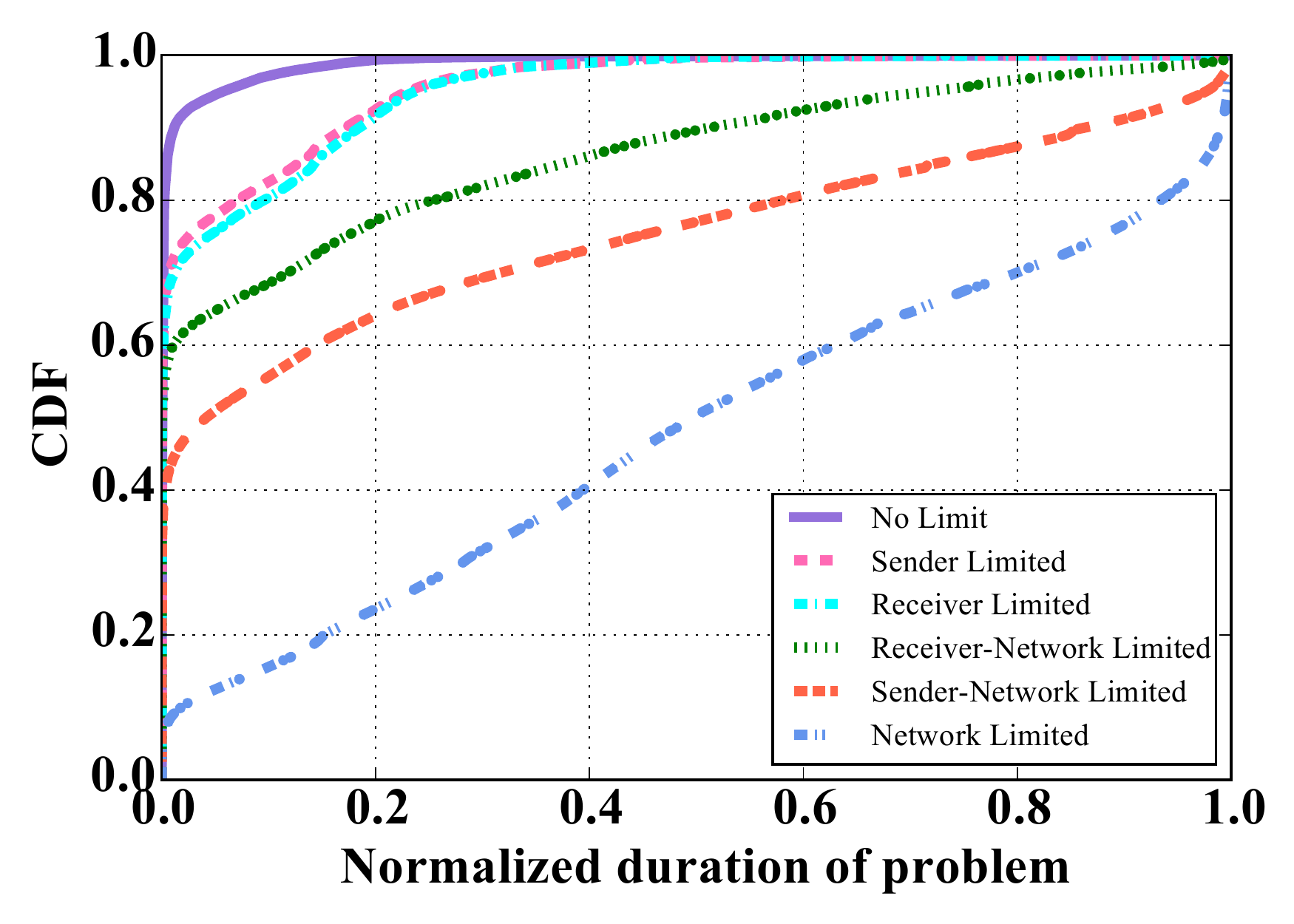}
\caption{Diagnosis Results for CAIDA traces}
\label{fig:bottlenecks}
\end{figure}

In Figure~\ref{fig:bottlenecks}, we present a CDF of the normalized fraction of time each class of performance problems limits a connection. Since the bottlenecks in a connection may change over time, we present the CDF of normalized duration of bottlenecks. We observe that: \emph{1.} 99\% of flows spend less than 1\% of their life in the no-limit state; this means that 99\% of flows have at least one bottleneck in more than 99\% of their lifetime. \emph{2.} Although sender and receiver problems have similar frequency and duration, we did not use the application reaction time to detect the sender-limited connections as the data is collected in core of the network as opposed to the edge, thus we expect the actual rate of sender-limited connections to be higher. \emph{3.} Many connections are bottlenecked by two factors simultaneously (e.g., sender-network or receiver-network). \emph{4.} About 90\% of the connections spend some time in the network-limited state, with almost half of them being network-limited 50\% of the time.~\footnote{Note that these results are from wide-area network and the characteristics of data-center connections are different from WAN.}

\subsection{Trade-offs in Accuracy and Overhead}
\label{s:two-stage-accuracy}
  Using the CAIDA trace, we evaluate the impact of our space optimizations on the accuracy of our measurement. To do this, we operate our software prototype in two modes: first, trades-off space for accuracy (emulate P4) and second, measures with high accuracy and precision (ground-truth). 

\begin{figure}[t!]
\centering
\includegraphics[height=2in]{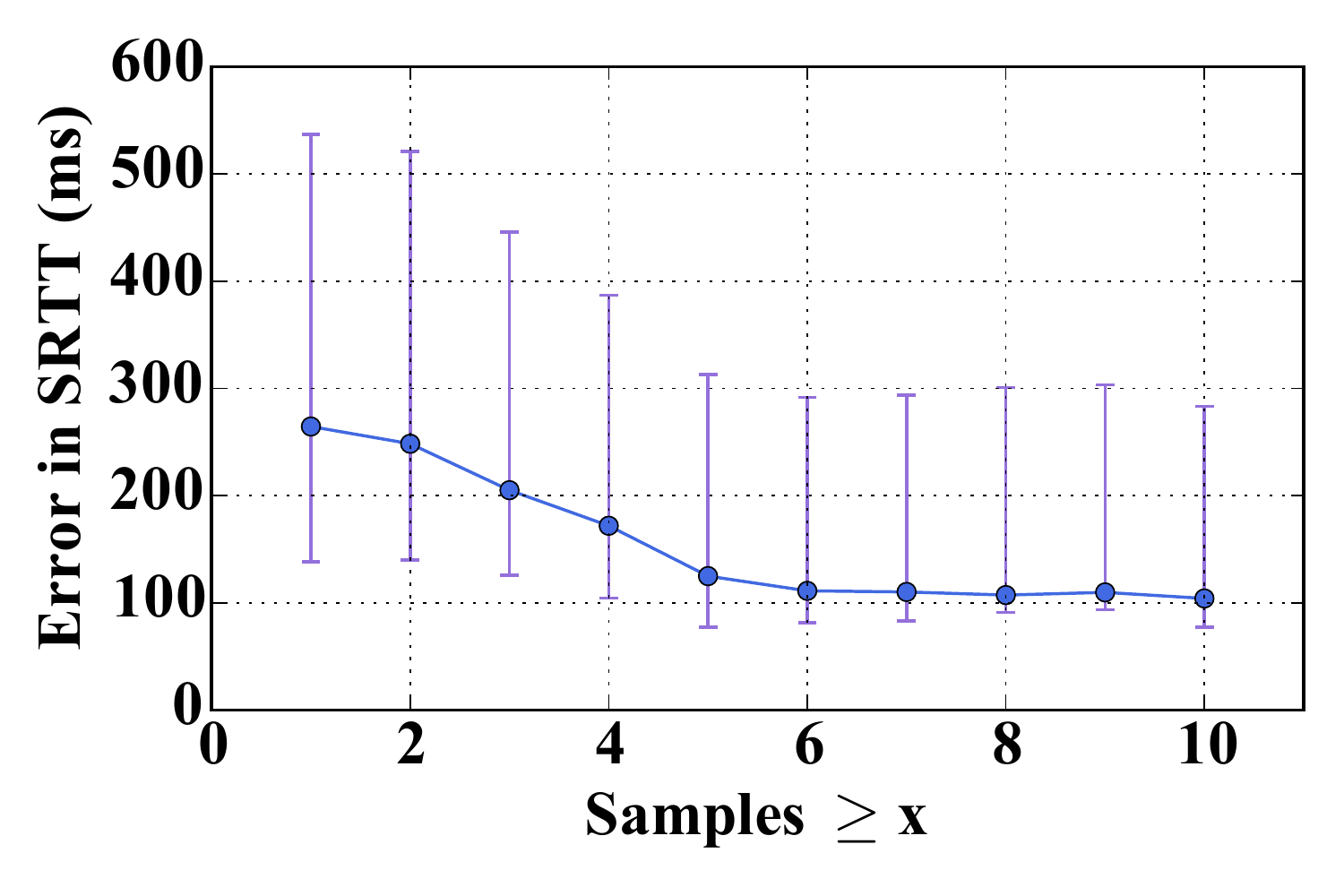}
\caption{Error in inferring SRTT with queue size of 1. As more packets are exchanged, the error decreases.}
\label{fig:srtt_error}
\end{figure}

\vspace{1mm}
\noindent\textbf{Limited queue size:} Our measurement shows that running \XYZ with an unbounded queue increases memory usage by about 9\%. In a network with higher bandwidth capacity, (i.e., high bandwidth delay product), more memory will be required. In Figure~\ref{fig:srtt_error}, we examine the error in SRTT when the queue size is bounded to one. We observe that when the queue size is bounded, \XYZ requires more samples to reduce error. Note these traces were collected in WAN, not a datacenter, hence the high RTTs. 

\vspace{1mm}

\begin{figure}[t!]
\centering
	\includegraphics[height=2in]{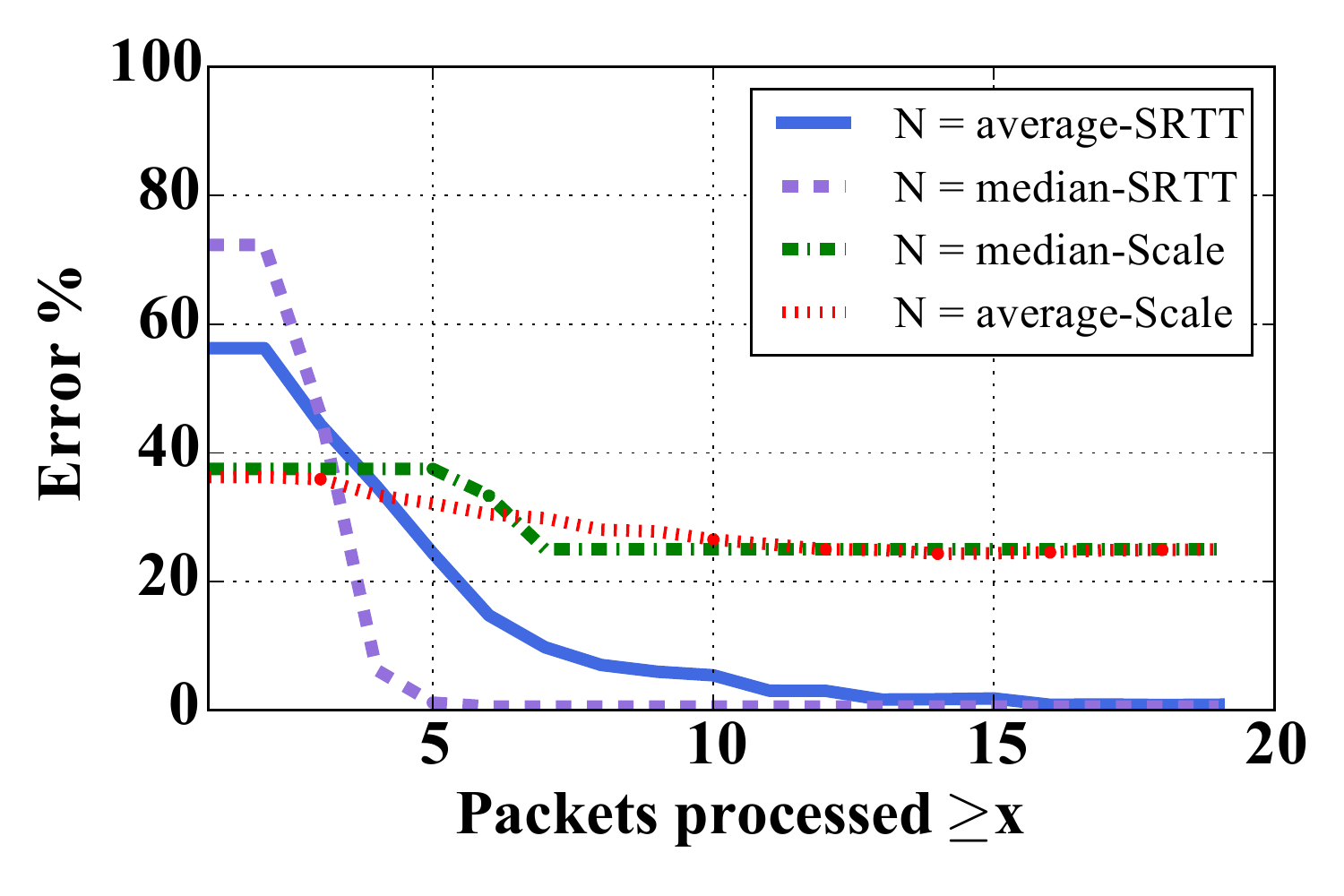}
\caption{Error in inferring options (MSS and wscale) midstream, the error rate decreases faster for MSS.}
\label{fig:mss_error}
\end{figure}

\noindent\textbf{Two-phase monitoring:}  Recall, two-phase monitoring offers a trade-off between accuracy of our heuristics and their memory overhead. We filter the connections to consider only those with known options, and use that as the ground-truth to compare against the midstream inferred constants. Figure~\ref{fig:mss_error} shows the error in inferring MSS and window scaling options according to Section~\ref{s:mid}.  While the error in MSS decreases as more packets are inspected, the inferred window scale value still differs from the ground-truth by about 20\%.  This is because most of these connections are not RWND-limited, hence the flight size values used in estimating the scale (Equation~\ref{eq:wscale}) do not approach the upper-bound imposed by the receiver. 

\section{Related Work}
\label{s:relatedwork}
Here, we discuss the differences of our approach with existing work in the following categories: 

\vspace{2mm}
\noindent\textbf{Offline packet trace analysis:}
Several tools analyze packet traces to find performance limitation~\cite{TRAT,Barford:2001:CPA:383576.383581,tcptrace} for a known TCP variant. Some~\cite{VND} store packet headers that facilitate diagnosis via running queries on headers. However, \emph{offline} analysis makes this category unsuitable for real-time diagnosis and introduces large data-collection overhead. 

\vspace{2mm}
\noindent\textbf{Measurement in the core:}
Some tools use coarse-grained metrics collected on switches, e.g., 5-minute SNMP counters~\cite{ Wu:2012:NAD:2377677.2377759}. Such metrics are not sufficiently fine-grained to diagnose the sources of poor performance~\cite{Bahl:2007:THR:1282380.1282383}; others~\cite{Liu:2016:MTC:2890955.2890964} focus on coordination between network switches to ensure the full packet stream is analyzed.  
Instead, we rely on the ``edge'' for better visibility and a simpler solution. 

\vspace{2mm}
\noindent\textbf{Instrumenting the network stack:}
Existing end-host techniques~\cite{wmust:2011,HONE,web10G,SNAP}, while appropriate for private clouds, are too invasive for a public IaaS cloud because they would run in the tenant's VM. 
\XYZ runs at the cloud-provider edge and infers the VM's internal state without tenant's cooperation.

\vspace{2mm}
\noindent\textbf{Tomography:}
Network tomography infers link-level properties (e.g., loss , delay) from end-to-end measurements~\cite{Caceres:1999,Nguyen:2005:BVA:2150193.2150204,832482,Nguyen:2007:NLI:1298306.1298339, 916661,Coates:2002:MLN:511334.511337} and may use linear~\cite{castro2004, 4016293} or Boolean algebra~\cite{4016293, Wei:2003:MID:948205.948220,4215834} to find congested links.
Our work differs from this body of work since we rely on \emph{direct and continuous measurement} of performance instead. Tomography can be complementary to \XYZ, e.g., to locate the congested link(s) among network-limited connections.  

\vspace{2mm}
\noindent\textbf{Machine learning:}
Recent work~\cite{Arzani:2016:TBG:2934872.2934884} makes the case for using machine learning (ML) in connection's performance diagnosis. Our solution differs in several aspects: 
first, instead on relying on ML, we leverage the fact that TCP is a well studied protocol and derive techniques based on common characteristics of TCP and its interaction with applications. Second, our solution is agnostic to application and TCP variant, while the ML would require new training for each variant. 

\vspace{2mm}
\noindent\textbf{Congestion control from the ``edge'':}
To enable new congestion control algorithms in multi-tenant clouds, \cite{Cronkite-Ratcliff:2016:VCC:2934872.2934889} offers virtualized congestion control in the hypervisor. Similarly, \cite{He:2016:ATV:2934872.2934903} enforces congestion control in the vswitch without cooperation from the tenant VM. 

\section{Conclusion}
\label{s:concl}
\XYZ helps cloud providers diagnose TCP connections without instrumenting the tenant VMs. Our solution can run at line rate in the hypervisor, NIC, or top-of-rack switch.
\bibliographystyle{abbrv}
\bibliography{ref}
\end{document}